\documentclass[12pt]{book}
\usepackage[dvips]{graphicx,color}
\usepackage{makeidx,tsukuba}
\usepackage{times}

\makeauthorindex
\makeindex

\begin{document}

\BookTitle{\itshape The 28th International Cosmic Ray Conference}
\CopyRight{\copyright 2003 by Universal Academy Press, Inc.}
\pagenumbering{arabic}

\chapter{Pierre Auger Atmosphere-Monitoring Lidar System}

\author{%
A.~Filip\v{c}i\v{c},$^1$ M.~Horvat,$^1$
\underline{D.~Veberi\v{c}},$^1$ D.~Zavrtanik,$^1$
M.~Zavrtanik,$^1$
M.~Chiosso,$^2$ R.~Mussa,$^2$ G.~Sequeiros,$^2$
M.A.~Mostafa,$^{2,3}$ and M.D.~Roberts$^3$
\\
(1) {\it Laboratory for Astroparticle Physics, Nova Gorica
Polytechnic, Slovenia}
\\
(2) {\it INFN--Torino, Italy}
\\
(3) {\it University of New Mexico, Albuquerque, USA}
}

\section*{Abstract}
The fluorescence-detection techniques of cosmic-ray air-shower
experiments require precise knowledge of atmospheric properties to
reconstruct air-shower energies. Up to now, the atmosphere in
desert-like areas was assumed to be stable enough so that occasional
calibration of atmospheric attenuation would suffice to reconstruct
shower profiles. However, serious difficulties have been reported in
recent fluorescence-detector experiments causing systematic errors in
cosmic ray spectra at extreme energies. Therefore, a scanning
backscatter lidar system has been constructed for the Pierre Auger
Observatory in Malarg\"ue, Argentina, where on-line atmospheric
monitoring will be performed. One lidar system is already deployed at
the Los Leones fluorescence detector (FD) site and the second one is
currently (April 2003) under construction at the Coihueco site. Next
to the established ones, a novel analysis method with assumption on
horizontal invariance, using multi-angle measurements is shown to
unambiguously measure optical depth, as well as absorption and
backscatter coefficient.

\section{Introduction}

The error in shower energy estimation is directly proportional to the
uncertainty in the optical depth between the fluorescence-light origin
(within the extensive air shower) and FD cameras [5]. Although
reasonable predictions can be obtained using atmospheric models (e.g.\
US Standard Atmosphere), they do not satisfactorily cover seasonal
variations nor occurrence of aerosol layers, typically accompanying
windy days and reaching up to 3\,km over the ground. As a calorimeter
of the FD, atmosphere thus requires on-line or at least periodic
monitoring of its optical properties. The lidar seems to be a
reasonable choice for this task and it is adopted not only by the
Pierre Auger project but apparently also by other cosmic-ray related
experiments [4]. In the following sections construction and analysis
methods used for the reconstruction of the lidar signal are presented.

\section{DAQ System}

The Pierre Auger lidar system is based on the BigSky Ultra, frequency
tripled Nd:YaG laser, which is able to transmit up to 20 pulses per
second, each with energy of 7\,mJ and 4\,ns duration (i.e.\ pulse
length 1.2\,m). The emitted wavelength of 355\,nm is in the
$300-400$\,nm range of the nitrogen fluorescence spectrum. The three
receiver telescopes were constructed using $80$\,cm diameter parabolic
mirrors with focal length of 41\,cm. The mirror is made of
aluminum-coated pyrex and protected with SiO$_2$. The backscattered
light is detected by a Hammamatsu R7400 photomultiplier with operating
voltage up to 1000\,V and gain up to $10^7$. To suppress background, a
broadband UG-1 filter with 60\% transmittance at 353\,nm and FWHM of
50\,nm is used. The distances between laser beam and the mirror
centers are $\sim1$\,m, and the entry point of the laser beam into the
telescope's field of view is at $\sim200$\,m. The whole system is
fully steerable with $0.1^\circ$ angular step. The signal is digitized
using a three-channel Licel transient recorder TR40-160 with 12\,bit
resolution at 40\,MHz sampling rate with 16k trace length combined
with 250\,MHz photon-counting system. Maximum detection range of the
hardware is thus, with this sampling rate and trace length, set to
60\,km. However, in optimal conditions atmospheric features only up to
30\,km are observed. In order to limit the huge dynamic range of the
lidar signal the laser is operated at 20\% of the maximal energy with
additional 10\% attenuation filter in the beam. Schematic view of the
whole DAQ system can be seen in Fig.~1--left. Typical lidar signal is
shown in Fig.~1--right. Photon counting channel of the digitizer is a
useful capability that can substantially extend the range for faint
signals and circumvent possible inaccuracy of the analog channel.

\begin{figure}[t]
\begin{center}
\includegraphics[height=10.5pc]{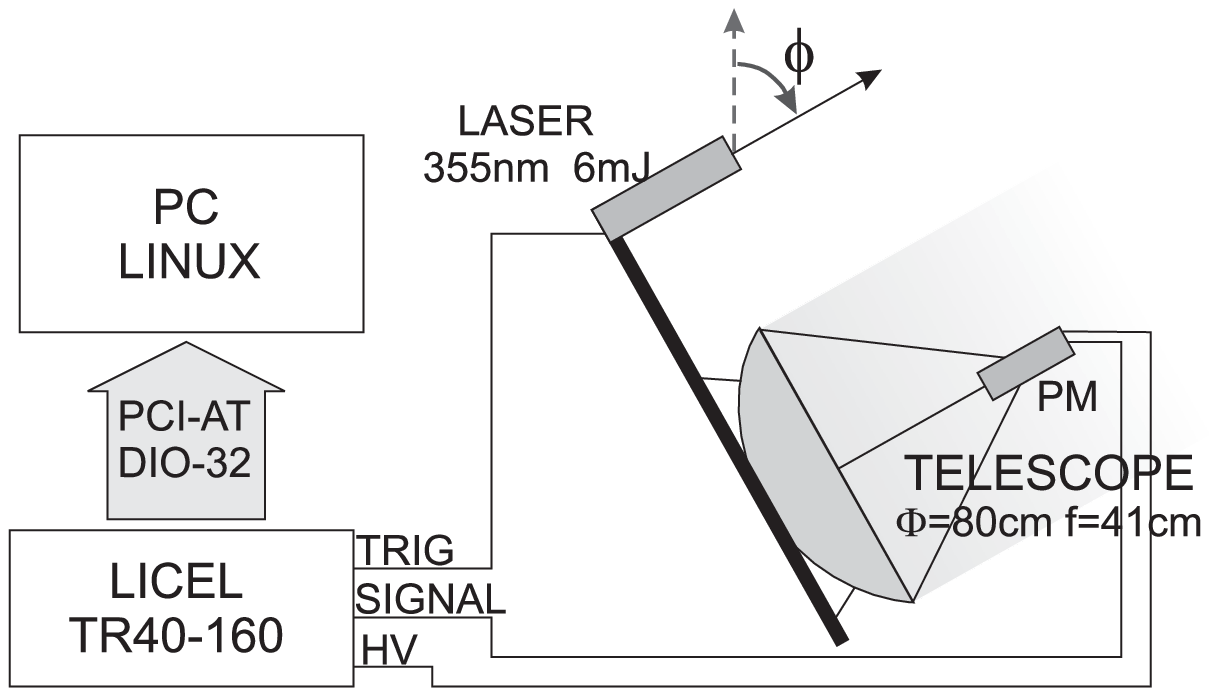}
\quad
\includegraphics[height=10.5pc]{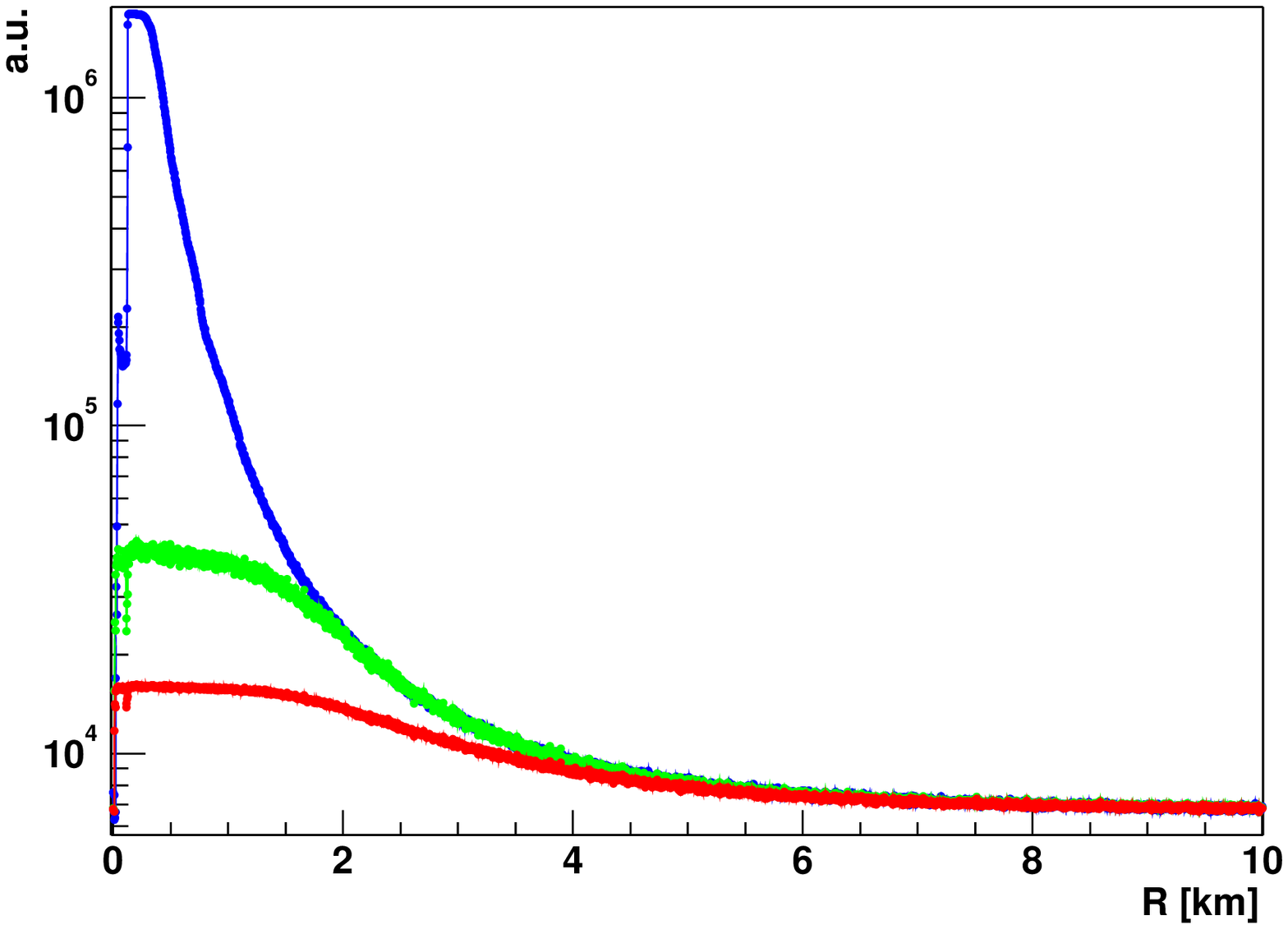}
\end{center}
\vspace{-0.5pc}
\caption{Schematic view of the lidar system (left). Example of
the typical lidar signal in a.u.\ as a function of range $R=ct/2$
(right). Top curve: analog channel, bottom: photon counting channel
(fitted to the analog channel within 5 and $10\,$km), and middle
curve: fitted photon channel with dead-time correction.}
\end{figure}

\section{Analysis}

So called \textit{lidar equation} [1,2,3], describing returned photon
flux, in fact represents an under-determined system of nonlinear
equations. Therefore, explicit solution of the equation does not
exist. In order to obtain some useful results, certain assumptions on
optical properties have to be made. One way is to postulate simple
potential expression relating the backscatter coefficient $\beta$ to
the attenuation (extinction) $\alpha$, as used by the Klett method
[1]. In the case of the Fernald method [2], both optical properties
are separated into molecular and aerosol part. Molecular part, i.e.\
Rayleigh scattering, is approximated with the assumed atmospheric
model, and the lidar equation is solved for the aerosol part.

In Fig.~2, Fernald method is applied on the representative measurement
taken with the Los Leones lidar station of the Pierre Auger
observatory. Integration of the attenuation $\alpha(h)$ results in
\textit{vertical optical depth} (VOD) $\tau$ that directly enters
estimation of the amount of fluorescence light [5], as measured by the
FD. Due to the aerosol layer near the ground (in Fig.~2--left reaching
up to 1.5\,km) the resulting VOD clearly differs from the predictions
of the (clean) atmosphere model (solid line). Judging from
Fig.~2--right, the difference $\Delta\tau$ between the atmospheric
model prediction and the result of the Fernald method can be as high
as several tenths of the unit. Neglecting such differences produces a
systematic underestimation of the shower energy (and correspondingly
the energy of primary particle), since in the first order $\Delta
E\propto\Delta\tau$.

\begin{figure}[t]
\begin{center}
\includegraphics[height=12.5pc]{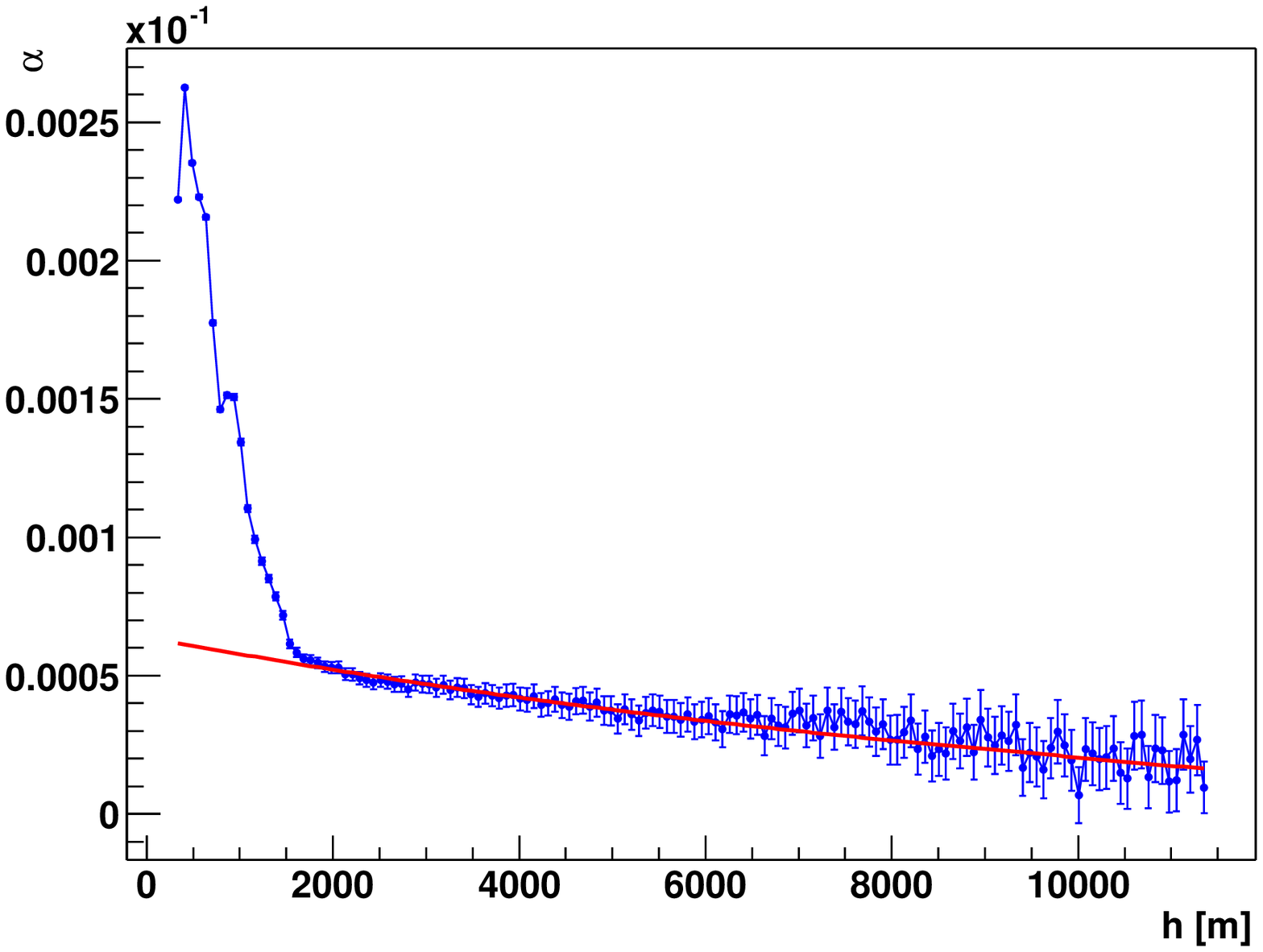}
\quad
\includegraphics[height=12.5pc]{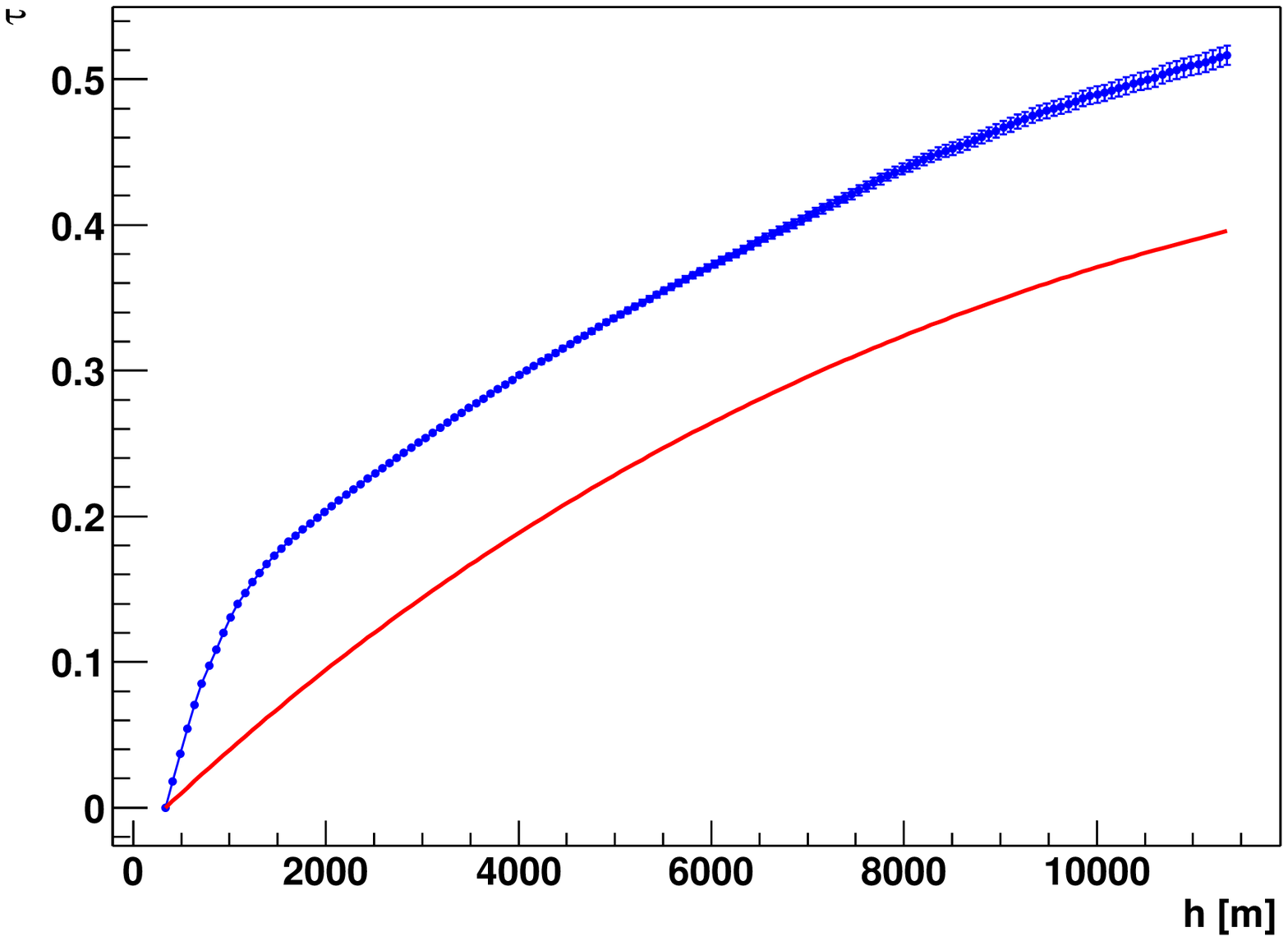}
\end{center}
\vspace{-0.5pc}
\caption{Total (molecular and aerosol) attenuation coefficient
$\alpha(h)$ obtained with Fernald inversion of a vertical shot of the
Los Leones lidar (left). Corresponding vertical optical depth
$\tau(h)$, (right). For comparison, attenuation and optical depth as
predicted by the US Standard Atmosphere (1976) model are drawn in
solid line.}
\end{figure}

Due to the fairly calm and stratified atmosphere above the huge plane
(Pampa Amarilla) where the Pierre Auger observatory is placed,
adequate assumption of the horizontal invariance can be made. Under
such an assumption the lidar equation is solved in a unique way with
the two- or multi-angle method [3] for both quantities, the
backscatter coefficient and the VOD measured relative to some
reference height ($h_0$, common in all lidar signals taken at
different angles). In Fig.~3, multi-angle reconstructions of
simultaneous lidar signals from two telescopes in the case of
relatively clear atmosphere is compared to the predictions of the
atmospheric model. Note, that the relative backscatter coefficient is
proportional to the relative atmospheric density, so that lidar system
can also serve as a monitoring tool for the atmospheric grammage,
important for the description of lateral shower development.

\begin{figure}[t]
\begin{center}
\includegraphics[height=12.1pc]{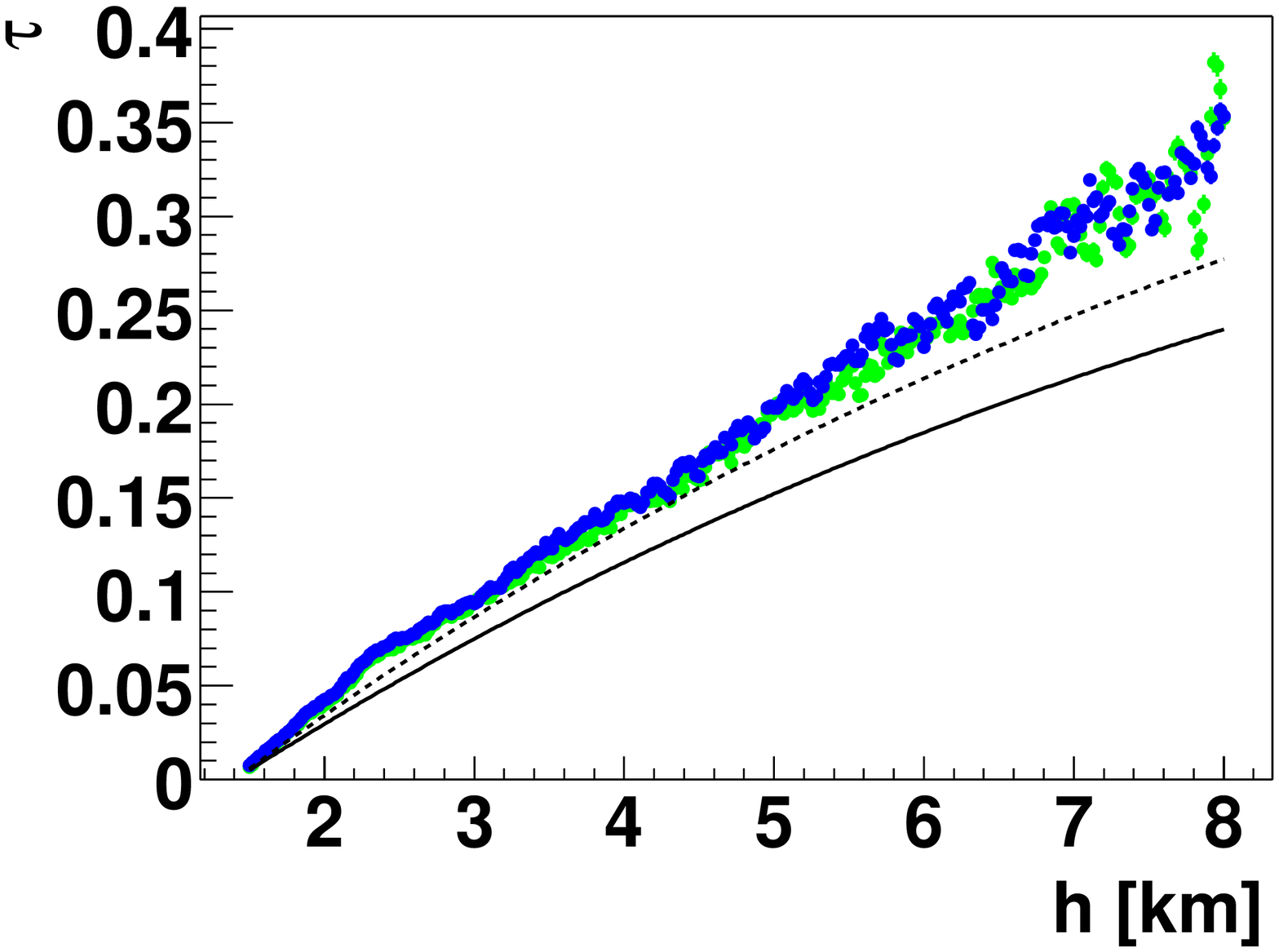}
\quad
\includegraphics[height=12.1pc]{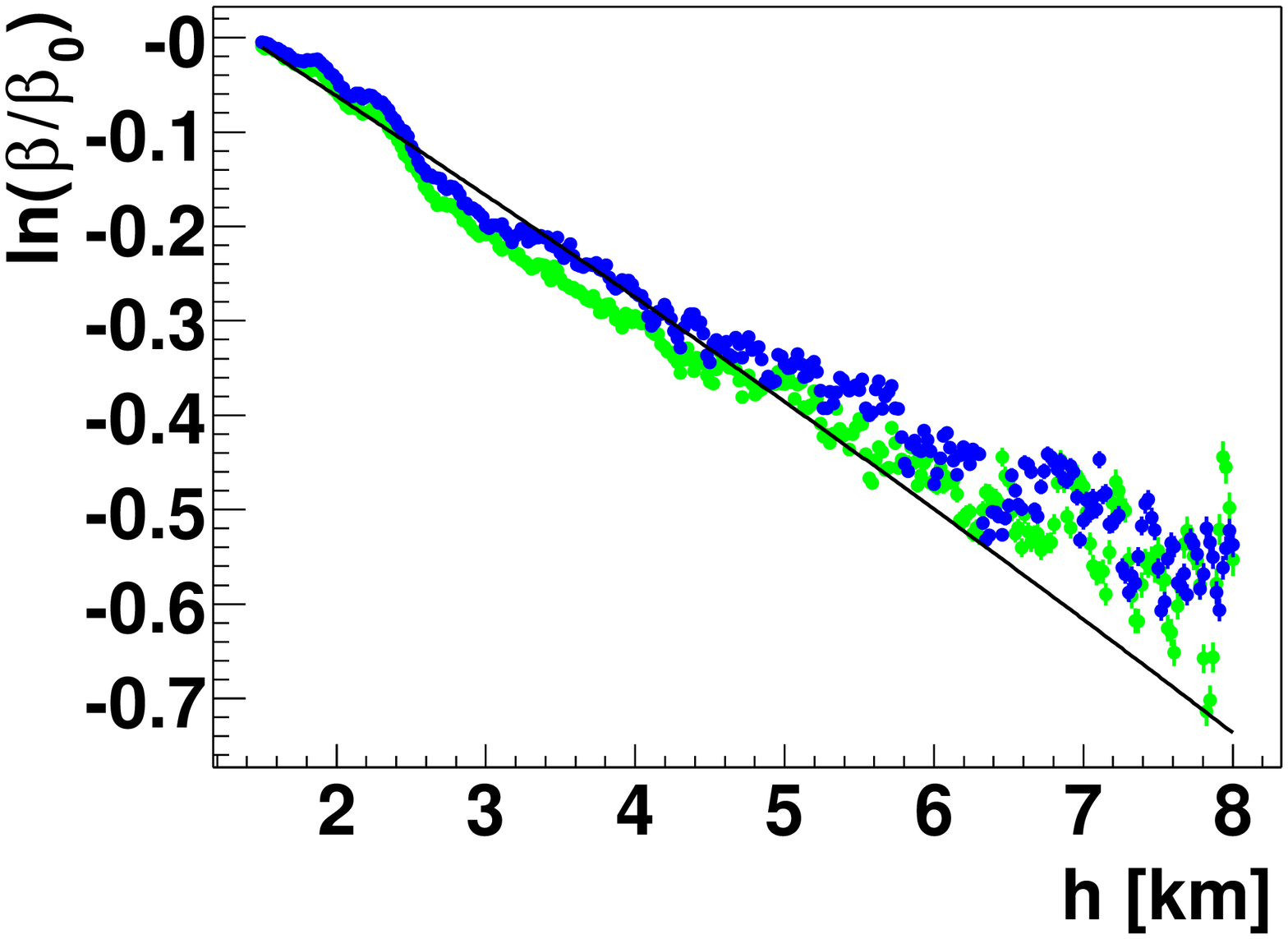}
\end{center}
\vspace{-0.5pc}
\caption{Vertical optical depth $\tau(h)$ (left) and relative
backscatter coefficient $\ln\beta(h)/\beta_0$ (right) obtained with
multi-angle analysis of lidar scans. Points are reconstructions of
simultaneous signals from two telescopes, solid line represents
prediction of the US Standard Atmosphere (1976) model, dashed: the
same model with depolarization effects included.}
\end{figure}

\section{Conclusion}

Aerosol layers near ground, occurrence of haze/clouds, and in a lesser
way seasonal fluctuations greatly influence shower energy estimation
as obtained by the FD measurements. Periodic monitoring of this
sensitive properties is thus unavoidable for any type of detection of
the fluorescence light originating from air showers in large
atmospheric volumes. Steerable lidar system can be successfully used
for such a demanding task. Nevertheless, careful selection,
optimization, and calibration of the corresponding lidar analysis
methods is strongly adverted.

\section*{References}
\re
1.\ Klett J.D.\ 1981, Appl.\ Opt.\ 20, 211
\re
2.\ Fernald F.G.\ 1984, Appl.\ Opt.\ 23, 652
\re
3.\ Filip\v{c}i\v{c} A.\ et al.\ 2003, Astropart.\ Phys.\ 18, 501
\re
4.\ Yamamoto T.\ et al.\ 2002, Nucl.\ Instr.\ and Meth.\ A 488, 191
\re
5.\ Argir\`o S.\ 2003, this Proceeding

\endofpaper
\end{document}